\documentclass[amsmath,amssymb,prl,showpacs,reprint, twocolumn]{revtex4-1}
\usepackage{graphicx}
\usepackage{dcolumn}
\usepackage{bm}

\usepackage{hyperref}

\usepackage[mathlines]{lineno}

\begin{document}

\preprint{APS/123-QED}

\title{Sensitivity to eV-scale Neutrinos of Experiments at a\\ Very Low Energy Neutrino Factory}
\author{J.H. Cobb}%
\author{C.D. Tunnell}
 \email{Corresponding author:  tunnell@fnal.gov}

\affiliation{Subdepartment of Particle Physics, University of Oxford, Denys Wilkinson Building,  
Keble Road, Oxford, OX1 3RH, UK}

\author{A.D. Bross}
\affiliation{
Fermilab, P.O. Box 500, Batavia, IL 60510-0500, USA
}

\date{\today}

\begin{abstract}
The results of LSND have yet to be confirmed at the $5 \sigma$-level.   An experiment is proposed utilizing a 3 GeV muon storage ring that would allow for both disappearance and appearance channels to be explored at short-baselines.  The appearance channel could provide well over $5 \sigma$ confirmation or rejection of the LSND result.  Other physics could also be performed at such a facility such as the measurement of electron-neutrino cross sections.  The sensitivity of experiments at a Very Low Energy Neutrino Factory (VLENF) to neutrinos at the eV-scale is presented.
\end{abstract}

\maketitle



Despite the experimental success of the past decade in establishing that neutrinos are massive and mix, whether they be produced in the Sun \cite{Aharmim:2008kc}, reactors \cite{Ahn:2006zza}, accelerators \cite{Adamson:2008zt}, or the atmosphere \cite{Wendell:2010md}, there exist nevertheless some anomalies.  All oscillation experiments to date can be explained with three neutrinos except LSND  --- and various experiments could be made to agree with either.  The LEP collider experiments \cite{lep} showed that additional light neutrinos cannot couple to the $Z$.  The MiniBooNe experiment has been unable to deny or confirm the LSND result since their data that uses the same anti-neutrino beam polarity as LSND agrees with the background-only hypothesis at 0.5\% and agrees with short-baseline oscillations at 8.7\%  \cite{PhysRevLett.105.181801}.   An experiment with a sensitivity greater than $5 \sigma$ is needed in order to refute or confirm evidence of neutrinos at the eV-scale.

The LSND data suggests a new mass splitting \cite{PhysRevLett.81.1774} that is potentially observed elsewhere.  Recently, a recalculation of reactor fluxes resulted in the \emph{reactor neutrino anomaly} \cite{Mueller:2011nm,Huber:2011wv} which, along with the \emph{Gallium anomaly} \cite{Acero:2007su}, are possibly caused by sterile neutrinos.  Global fits reconcile these anomalies into a framework that introduces a sterile neutrino at the eV-scale, \emph{e.g.}, \cite{Giunti:2011gz,Kopp:2011qd}.  Additional indications come from cosmology where WMAP favors more than three neutrinos \cite{Komatsu:2010fb}.

The Very Low Energy Neutrino Factory (VLENF) would utilize a muon storage ring to  study eV-scale oscillation physics and measure cross sections (including $\nu_e$).   Pions are collected from a target and injected into a storage ring where they decay to muons.  The storage ring is optimized for 2 GeV muons where the energy is optimized for the needs of both oscillation and cross section physics.  The muons decay according to $\mu^+ \to e^+ \bar{\nu}_\mu \nu_e$.  Straight sections in the storage ring result in neutrinos directed at a near- and far-detector. 

The storage ring could be a fixed field alternating gradient (FFAG) lattice for large momentum acceptance, which is important given the large momentum spectrum after the target.  The pricing and engineering of FFAGs are well-understood because of experience building them \cite{c:ffag}.  The design would require only normal conducting magnets which simplifies construction, commissioning, and operations.    Design work for the injection into the storage ring and the particle collection downstream of the target are underway.  

 The near detector will be placed at 20-50 meters from the end of the straight and will measure neutrino-nucleon cross sections of interest to future long-baseline experiments including the first precision measurement of $\nu_e$ cross sections.  The far detector at 800 m would measure disappearance and wrong-sign muon appearance channels.  The detector would need to be magnetized for the wrong-sign muon appearance channel.   Numerous possibilities exist for detector technologies that include liquid argon, MINOS inspired, and totally active scintillating detectors.  For the purposes of this study, a detector inspired by MINOS but with thinner plates is assumed.  The experiment will take advantage of the ``golden channel" of oscillation appearance $\nu_e \to \nu_\mu$ where the resulting final state is a wrong-sign muon.  

The probability $\nu_e \to \nu_\mu$ depends on the mixing matrix, $U$.  Let $R_{ij}$ be a rotation between the $i$th and $j$th mass eigenstates without CP violation. For $N$ neutrinos, $R_{ij}$ has dimension $N \times N$. By convention, the three neutrino mixing matrix is $U_\text{PMNS} = R_{23} R_{13} R_{12}$. In the (3+1) model of neutrino oscillations, extra rotations can be introduced such that the mixing matrix is $U_\text{(3+1)} = R_{34} R_{24} R_{14} U_\text{PMNS}$.  Given that $\Delta m^2_{41} >> \Delta m^2_{31}$, $U_\text{PMNS}$ can be approximated by the identity matrix (\emph{ie.} the ``short-baseline approximation").  It then follows that $| U_{e 4}|^2 = \sin(\theta_{14})$ and $ | U_{\mu 4}|^2 = \sin(\theta_{24}) \cos(\theta_{14}) $.

The oscillation probabilities for appearance and disappearance, respectively, are:
\begin{eqnarray}
\label{eq:prob}\text{P}_{\nu_e \to \nu_\mu} =& 4 | U_{e 4}|^2 |U_{\mu 4}|^2 \sin^2 \left(\frac{\Delta m^2_{41} L}{4 E}\right),\\
\text{P}_{\nu_\alpha \to \nu_\alpha} =& 1 - \left[4 |U_{\alpha 4}|^2 (1 - |U_{\alpha 4}|^2)\right] \sin^2 \left(\frac{\Delta m^2_{41} L}{4 E}\right).
\end{eqnarray}

\noindent
Disappearance measurements will constrain $| U_{\mu 4}|^2$ and $| U_{e 4}|^2$ while the appearance channel measures their product $|U_{e 4}|^2 |U_{\mu 4}|^2$.   These parameters are over-constrained.  Only the appearance channel will be assumed since the physics sensitivity of disappearance measurements is well-understood for short-baseline experiments.  Without oscillations there would be $\sim10^5$ charge current (CC) events in the far detector.  The disappearance oscillation probability, if LSND is correct, would be on the order of a few percent for this experiment; the statistical errors and systematic errors are expected to be about half a percent from the presence of a near detector and beam instrumentation.

The oscillation sensitivities have been computed using the GLoBES software (version 3.1.10)  \cite{Huber:2004ka,Huber:2007ji}.  Since GLoBES, by default, only allows for a $3 \times 3$ mixing matrix, the SNU (version 1.1) add-on  \cite{Kopp:2006wp,Kopp:2007ne} is used to extend computations in GLoBES to $4\times4$ mixing matrices.

Unlike previous analyses,  the far detector approximation of the source and detector being treated as point-sources cannot be used since the size of the detector and accelerator straight are comparable to the baseline.  This study computes the neutrino flux by integrating the phase space of the stored muons using Monte Carlo (MC) integration.  The beam occupies a 6D phase space ($x$, $y$, $z$, $p_x$, $p_y$, $p_z$) and the detector has a $6\text{ m} \times 6\text{ m}$ cross section.  A random point is chosen within the beam phase space and within the detector volume.  The transverse phase space is represented by the Twiss parameters $\alpha = 0$ and $\beta = 25 \text{ m}$ where the $1 \sigma$ Gaussian emittance is assumed to be $15 \text{ mm}$.  It follows that the spread in, for example, $x$ is $\sigma_x = \sqrt{\beta \epsilon}$ and the angular divergence in $x$ is $\sigma_{x'} = \sqrt{\epsilon / \beta}$.  The longitudinal phase space ($z$ and $p_z$) is described by assuming a uniform distribution in $z$ and $p_z = (2 \pm 20\%) \text{ GeV}$. The code for the analysis herein is available online \cite{vlenf_tools} under the GPL license \cite{gpl}.  

This analysis assumes $2 \times 10^{17}$ decays of $\mu^+$ in each straight, normalized to an exposure similar to MiniBooNe of $10^{21}$ protons on target (POT).  The far detector at 800 m consists of a kilotonne of fiducial target mass.  The background rejection for the wrong-sign muons is assumed to be $10^{-4}$ for the $\bar{\nu}_\mu \to \bar{\nu}_\mu$ neutral current (NC) events and $10^{-5}$ for the charge misidentification of $\bar{\nu}_\mu \to \bar{\nu}_{\mu}$ CC events.  The backgrounds from $\nu_e \to \nu_e$ CC and NC are negligible. The $\nu_e \to \nu_\mu$ CC events have a 90\% detection efficiency.  These efficiencies were extrapolated from previous studies \cite{laing} and reconfirmation of these calculations for energies below 2 GeV is still required.  

\begin{figure}[t]	
\begin{center}
\includegraphics[width=1.0\columnwidth]{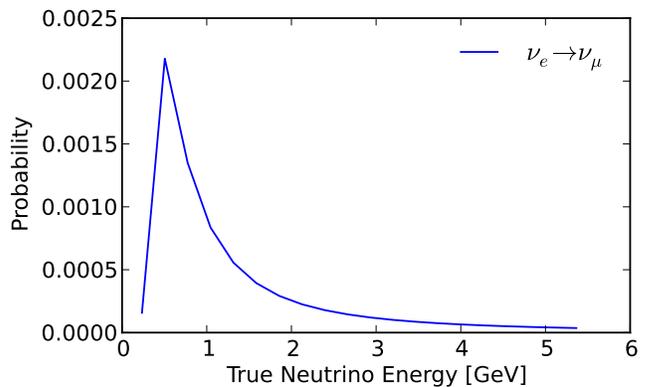}
\end{center}
\vspace*{-1cm}
\caption{\label{fig:osc_prob}The oscillation probability for the ``golden channel" $\nu_e \to \nu_\mu$ using the (3+1) oscillation parameters in TABLE \ref{tab:params}.  A baseline of 800 meters is assumed.  This numerical treatment of the binned oscillation probability agrees with Eq.~\ref{eq:prob}.}
\end{figure}

To initially gauge the sensitivity of this experimental setup, some oscillation parameters must be assumed that include the LSND effect.  The best-fit data for (3+1) sterile neutrinos is used \cite{Giunti:2011hn} where the MB $\bar{\nu}$ and LSND $\bar{\nu}$ data was fit (TABLE~\ref{tab:params}).   Lower energy neutrinos contain the most information about the mixing matrix (FIG.~\ref{fig:osc_prob}).

\begin{table}[b]
\caption{\label{tab:params}%
Best-fit oscillation parameters for the (3+1) sterile neutrino scenario for MB $\bar{\nu}$ and LSND $\bar{\nu}$ data \cite{Giunti:2011hn}.}
\begin{ruledtabular}
\begin{tabular}{lr}
\textrm{Parameter}&
\textrm{Value}\\
\colrule
$\Delta m^2_{41}$ [$\text{eV}^2$] & 0.89\\
$|U_{e4}|^2$ & 0.025\\
$|U_{\mu 4}|^2$ & 0.023\\
\end{tabular}
\end{ruledtabular}
\end{table}

\begin{figure}[t]
\begin{center}
\includegraphics[width=\columnwidth]{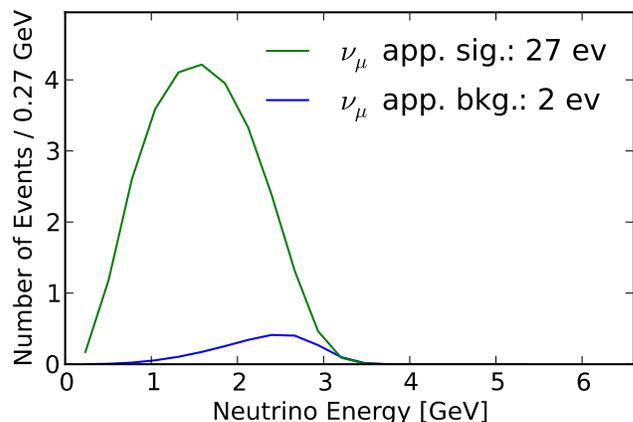}
\end{center}
\vspace*{-1cm}
\caption{The spectrum of signal and background events at the far-detector.   The number of wrong-sign muon appearance signal events is ~27 which arises through the ``golden channel" oscillation probability $\nu_e \to \nu_\mu$.  Roughly 2 background events are in the signal box and correspond to both $\bar{\nu}_\mu$ CC with charge misidentification and $\bar{\nu}_\mu$ NC events where a short-lived pion has faked a wrong-sign muon.  An exposure of $10^{21}$ POT is assumed.}
\label{fig:ws_spect}
\end{figure}

  The best-fit values allow for an estimation of the event rates.  The true number of events without efficiencies is shown in TABLE~\ref{tab:table1}.   Applying the assumed efficiencies of $10^{-5}$ and $10^{-4}$ for $\bar{\nu}_\mu \to \bar{\nu}_\mu$ CC and $\bar{\nu}_X \to \bar{\nu}_X$ NC, respectively, reveals the event spectrum (FIG.~\ref{fig:ws_spect}).   After cuts, there are ~27 signal events and ~2 background events.  

\begin{table}[b]
\caption{\label{tab:table1}%
A table of the raw event rates. The first row corresponds to the ``golden channel" appearance signal. The other rows are potential backgrounds to the signal.  The background that drives the analysis is $\bar{\nu}_\mu \to \bar{\nu}_\mu$.}
\begin{ruledtabular}
\begin{tabular}{lcr}
\textrm{Channel}&
\textrm{Interaction}&
\textrm{Pre-cuts}\\
\colrule
$\nu_e \to \nu_\mu$ & CC & 30 \\
\hline
$\bar{\nu}_X \to \bar{\nu}_X$ & NC &16850\\
$\bar{\nu}_\mu \to \bar{\nu}_\mu$ & CC &42545\\
$\nu_e \to \nu_e$ & CC &78974 \\
\end{tabular}
\end{ruledtabular}
\end{table}
       	    	     	     
\begin{figure}[t]	
\begin{center}
\includegraphics[width=1.0\columnwidth]{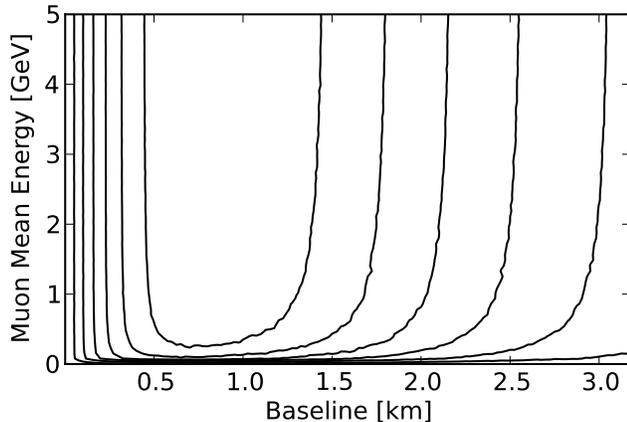}
\end{center}
\vspace*{-1cm}
\caption{\label{fig:l_v_e}Optimization of the mean stored muon energy and far detector baseline. Above muon energies of 1 GeV, appearance physics is not sensitive to the muon energy.  For a fixed neutrino energy, the boosted muon frame will result in more events the more it is boosted. The optimal baseline is around 800 meters.}
\end{figure}

In practice the muon energy is fixed by the needs of the cross section physics.  However, given a fixed fiducial mass, the baseline and muon energy can be optimized for short-baseline physics.  The metric for comparison is the $\chi^2$ between simulations for the (3+1) best-fit mentioned earlier and the background-only hypothesis. The short-baseline oscillation physics reach is comparable at energies above 2 GeV and the optimal baseline is around 800 meters (FIG. \ref{fig:l_v_e}).  

Having chosen a baseline and energy, it is possible to investigate the physics reach of this facility if we appropriately define the desired $\chi^2$ value.  The design requirement is to measure the LSND effect at ``$5 \sigma$" but that can be ambiguously defined.  Herein ``$n \sigma$" is defined according to the Gaussian probability of the respective deviation.  Let $\text{CDF}_X$ be the cumulative distribution function of some distribution $X$.  If $G$ is a Gaussian with $\mu = 0$ and $\sigma = 1$, then the $p$-value of a $5 \sigma$ effect is $1 - \text{CDF}_G(5) \simeq 3 \times 10^{-7}$.  Similarly, if $\chi^2_2$ is the chi-squared distribution with two degrees of freedom, then the required value of the $\chi^2$ to be an $n \sigma$ effect is $\text{CDF}^{-1}_{\chi^2_2}(\text{CDF}_G(n))$.


\begin{figure}[t]
\begin{center}
\includegraphics[width=\columnwidth]{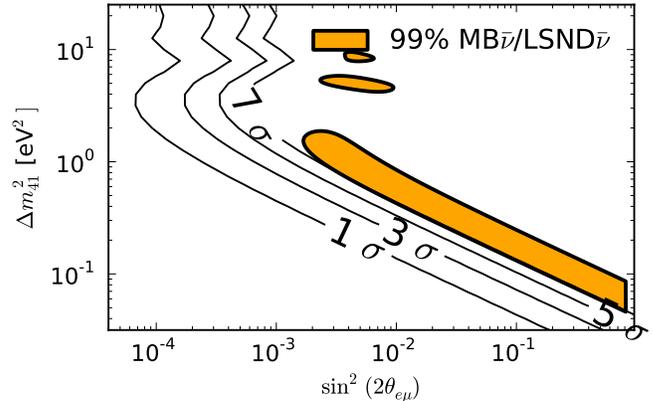}
\end{center}
\vspace*{-1cm}
\caption{\label{fig:ffag_sens}Sensitivity to sterile parameters.  The orange band corresponds to the LSND $\bar{\nu}$ and MiniBooNe $\bar{\nu}$ sterile neutrino fit performed by Giunti and Laveder \cite{Giunti:2011gz} where they assumed only one extra massive neutrino.  The contours are the sensitivities of an experiment at a 3 GeV FFAG storage ring using parameters defined in the text. The entire 99\% confidence interval for the LSND $\bar{\nu}$ and MiniBooNe $\bar{\nu}$ results would be confirmed or excluded at $7 \sigma$.}
\end{figure}

The $\chi^2$ is computed using the pull-method, which allows for systematics to be included.  The signal and background normalization errors are assigned to be 2\% and 20\%, respectively, and these systematic uncertainties are marginalized over.  Spectral information is also used.  The sensitivity to new physics is shown in FIG.~\ref{fig:ffag_sens}.  The 99\% confidence interval of a fit in a (3+1) scheme to the MiniBooNe $\bar{\nu}$ and LSND $\bar{\nu}$ data is shown and excluded at $7 \sigma$.  This meets and exceeds the requested criterion for future eV-scale experiments.

Some effects are ignored that could reduce the significance of the result.  For example, not included are backgrounds associated with cosmic muons.  This could be problematic since muons fill the storage ring which results in a large duty factor.  One solution, if it is an issue, would be to put RF cavities in the storage ring to bunch the beam. This merits further study.

It has been demonstrated that a VLENF has sensitivity to eV-scale oscillations.  Further optimizations of the target horn, collection, injection, and the dynamic aperture of the storage ring will increase the neutrino flux.  But previously unthought-of backgrounds may arise and detector charge identification performance is difficult at 1 GeV and below.  A design study must be performed to ensure enough contingency in the sensitivity to guarantee a firm confirmation or refutation of the LSND effect.  Further study is required.

\emph{Thanks} for guidance and useful discussions with Joachim Kopp, David Neuffer, Kenneth Long, and Alain Blondel.  Also thanks to Nick Ryder for a careful reading.

\bibliography{vlenf_sterile}

\providecommand{\noopsort}[1]{}\providecommand{\singleletter}[1]{#1}%
\begin{thebibliography}{22}%
\makeatletter
\providecommand \@ifxundefined [1]{%
 \@ifx{#1\undefined}
}%
\providecommand \@ifnum [1]{%
 \ifnum #1\expandafter \@firstoftwo
 \else \expandafter \@secondoftwo
 \fi
}%
\providecommand \@ifx [1]{%
 \ifx #1\expandafter \@firstoftwo
 \else \expandafter \@secondoftwo
 \fi
}%
\providecommand \natexlab [1]{#1}%
\providecommand \enquote  [1]{``#1''}%
\providecommand \bibnamefont  [1]{#1}%
\providecommand \bibfnamefont [1]{#1}%
\providecommand \citenamefont [1]{#1}%
\providecommand \href@noop [0]{\@secondoftwo}%
\providecommand \href [0]{\begingroup \@sanitize@url \@href}%
\providecommand \@href[1]{\@@startlink{#1}\@@href}%
\providecommand \@@href[1]{\endgroup#1\@@endlink}%
\providecommand \@sanitize@url [0]{\catcode `\\12\catcode `\$12\catcode
  `\&12\catcode `\#12\catcode `\^12\catcode `\_12\catcode `\%12\relax}%
\providecommand \@@startlink[1]{}%
\providecommand \@@endlink[0]{}%
\providecommand \url  [0]{\begingroup\@sanitize@url \@url }%
\providecommand \@url [1]{\endgroup\@href {#1}{\urlprefix }}%
\providecommand \urlprefix  [0]{URL }%
\providecommand \Eprint [0]{\href }%
\providecommand \doibase [0]{http://dx.doi.org/}%
\providecommand \selectlanguage [0]{\@gobble}%
\providecommand \bibinfo  [0]{\@secondoftwo}%
\providecommand \bibfield  [0]{\@secondoftwo}%
\providecommand \translation [1]{[#1]}%
\providecommand \BibitemOpen [0]{}%
\providecommand \bibitemStop [0]{}%
\providecommand \bibitemNoStop [0]{.\EOS\space}%
\providecommand \EOS [0]{\spacefactor3000\relax}%
\providecommand \BibitemShut  [1]{\csname bibitem#1\endcsname}%
\let\auto@bib@innerbib\@empty
\bibitem [{\citenamefont {Aharmim}\ \emph {et~al.}(2008)\citenamefont {Aharmim}
  \emph {et~al.}}]{Aharmim:2008kc}%
  \BibitemOpen
  \bibfield  {author} {\bibinfo {author} {\bibfnamefont {B.}~\bibnamefont
  {Aharmim}} \emph {et~al.} (\bibinfo {collaboration} {SNO}),\ }\href {\doibase
  10.1103/PhysRevLett.101.111301} {\bibfield  {journal} {\bibinfo  {journal}
  {Phys. Rev. Lett.}\ }\textbf {\bibinfo {volume} {101}},\ \bibinfo {pages}
  {111301} (\bibinfo {year} {2008})},\ \Eprint {http://arxiv.org/abs/0806.0989}
  {arXiv:0806.0989 [nucl-ex]} \BibitemShut {NoStop}%
\bibitem [{\citenamefont {Ahn}\ \emph {et~al.}(2006)\citenamefont {Ahn} \emph
  {et~al.}}]{Ahn:2006zza}%
  \BibitemOpen
  \bibfield  {author} {\bibinfo {author} {\bibfnamefont {M.~H.}\ \bibnamefont
  {Ahn}} \emph {et~al.} (\bibinfo {collaboration} {K2K}),\ }\href {\doibase
  10.1103/PhysRevD.74.072003} {\bibfield  {journal} {\bibinfo  {journal} {Phys.
  Rev.}\ }\textbf {\bibinfo {volume} {D74}},\ \bibinfo {pages} {072003}
  (\bibinfo {year} {2006})},\ \Eprint {http://arxiv.org/abs/hep-ex/0606032}
  {arXiv:hep-ex/0606032} \BibitemShut {NoStop}%
\bibitem [{\citenamefont {Adamson}\ \emph {et~al.}(2008)\citenamefont {Adamson}
  \emph {et~al.}}]{Adamson:2008zt}%
  \BibitemOpen
  \bibfield  {author} {\bibinfo {author} {\bibfnamefont {P.}~\bibnamefont
  {Adamson}} \emph {et~al.} (\bibinfo {collaboration} {MINOS}),\ }\href
  {\doibase 10.1103/PhysRevLett.101.131802} {\bibfield  {journal} {\bibinfo
  {journal} {Phys. Rev. Lett.}\ }\textbf {\bibinfo {volume} {101}},\ \bibinfo
  {pages} {131802} (\bibinfo {year} {2008})},\ \Eprint
  {http://arxiv.org/abs/0806.2237} {arXiv:0806.2237 [hep-ex]} \BibitemShut
  {NoStop}%
\bibitem [{\citenamefont {Wendell}\ \emph {et~al.}(2010)\citenamefont {Wendell}
  \emph {et~al.}}]{Wendell:2010md}%
  \BibitemOpen
  \bibfield  {author} {\bibinfo {author} {\bibfnamefont {R.}~\bibnamefont
  {Wendell}} \emph {et~al.} (\bibinfo {collaboration} {Kamiokande}),\ }\href
  {\doibase 10.1103/PhysRevD.81.092004} {\bibfield  {journal} {\bibinfo
  {journal} {Phys. Rev.}\ }\textbf {\bibinfo {volume} {D81}},\ \bibinfo {pages}
  {092004} (\bibinfo {year} {2010})},\ \Eprint {http://arxiv.org/abs/1002.3471}
  {arXiv:1002.3471 [hep-ex]} \BibitemShut {NoStop}%
\bibitem [{lep(2006)}]{lep}%
  \BibitemOpen
  \href {\doibase 10.1016/j.physrep.2005.12.006} {\bibfield  {journal}
  {\bibinfo  {journal} {Phys. Rept.}\ }\textbf {\bibinfo {volume} {427}},\
  \bibinfo {pages} {257} (\bibinfo {year} {2006})},\ \Eprint
  {http://arxiv.org/abs/hep-ex/0509008} {arXiv:hep-ex/0509008} \BibitemShut
  {NoStop}%
\bibitem [{\citenamefont {Aguilar-Arevalo}\ \emph {et~al.}(2010)\citenamefont
  {Aguilar-Arevalo} \emph {et~al.}}]{PhysRevLett.105.181801}%
  \BibitemOpen
  \bibfield  {author} {\bibinfo {author} {\bibfnamefont {A.~A.}\ \bibnamefont
  {Aguilar-Arevalo}} \emph {et~al.} (\bibinfo {collaboration} {MiniBooNE
  Collaboration}),\ }\href {\doibase 10.1103/PhysRevLett.105.181801} {\bibfield
   {journal} {\bibinfo  {journal} {Phys. Rev. Lett.}\ }\textbf {\bibinfo
  {volume} {105}},\ \bibinfo {pages} {181801} (\bibinfo {year}
  {2010})}\BibitemShut {NoStop}%
\bibitem [{\citenamefont {Athanassopoulos}\ \emph {et~al.}(1998)\citenamefont
  {Athanassopoulos} \emph {et~al.}}]{PhysRevLett.81.1774}%
  \BibitemOpen
  \bibfield  {author} {\bibinfo {author} {\bibfnamefont {C.}~\bibnamefont
  {Athanassopoulos}} \emph {et~al.} (\bibinfo {collaboration} {LSND
  Collaboration}),\ }\href {\doibase 10.1103/PhysRevLett.81.1774} {\bibfield
  {journal} {\bibinfo  {journal} {Phys. Rev. Lett.}\ }\textbf {\bibinfo
  {volume} {81}},\ \bibinfo {pages} {1774} (\bibinfo {year}
  {1998})}\BibitemShut {NoStop}%
\bibitem [{\citenamefont {Mueller}\ \emph {et~al.}(2011)\citenamefont {Mueller}
  \emph {et~al.}}]{Mueller:2011nm}%
  \BibitemOpen
  \bibfield  {author} {\bibinfo {author} {\bibfnamefont {T.~A.}\ \bibnamefont
  {Mueller}} \emph {et~al.},\ }\href {\doibase 10.1103/PhysRevC.83.054615}
  {\bibfield  {journal} {\bibinfo  {journal} {Phys. Rev.}\ }\textbf {\bibinfo
  {volume} {C83}},\ \bibinfo {pages} {054615} (\bibinfo {year} {2011})},\
  \Eprint {http://arxiv.org/abs/1101.2663} {arXiv:1101.2663 [hep-ex]}
  \BibitemShut {NoStop}%
\bibitem [{\citenamefont {Huber}(2011)}]{Huber:2011wv}%
  \BibitemOpen
  \bibfield  {author} {\bibinfo {author} {\bibfnamefont {P.}~\bibnamefont
  {Huber}},\ }\href {\doibase 10.1103/PhysRevC.84.024617} {\bibfield  {journal}
  {\bibinfo  {journal} {Phys. Rev.}\ }\textbf {\bibinfo {volume} {C84}},\
  \bibinfo {pages} {024617} (\bibinfo {year} {2011})},\ \Eprint
  {http://arxiv.org/abs/1106.0687} {arXiv:1106.0687 [hep-ph]} \BibitemShut
  {NoStop}%
\bibitem [{\citenamefont {Acero}\ \emph {et~al.}(2008)\citenamefont {Acero},
  \citenamefont {Giunti},\ and\ \citenamefont {Laveder}}]{Acero:2007su}%
  \BibitemOpen
  \bibfield  {author} {\bibinfo {author} {\bibfnamefont {M.~A.}\ \bibnamefont
  {Acero}}, \bibinfo {author} {\bibfnamefont {C.}~\bibnamefont {Giunti}}, \
  and\ \bibinfo {author} {\bibfnamefont {M.}~\bibnamefont {Laveder}},\ }\href
  {\doibase 10.1103/PhysRevD.78.073009} {\bibfield  {journal} {\bibinfo
  {journal} {Phys. Rev.}\ }\textbf {\bibinfo {volume} {D78}},\ \bibinfo {pages}
  {073009} (\bibinfo {year} {2008})},\ \Eprint {http://arxiv.org/abs/0711.4222}
  {arXiv:0711.4222 [hep-ph]} \BibitemShut {NoStop}%
\bibitem [{\citenamefont {Giunti}\ and\ \citenamefont
  {Laveder}(2011{\natexlab{a}})}]{Giunti:2011gz}%
  \BibitemOpen
  \bibfield  {author} {\bibinfo {author} {\bibfnamefont {C.}~\bibnamefont
  {Giunti}}\ and\ \bibinfo {author} {\bibfnamefont {M.}~\bibnamefont
  {Laveder}},\ }\href@noop {} {\  (\bibinfo {year} {2011}{\natexlab{a}})},\
  \Eprint {http://arxiv.org/abs/1107.1452} {arXiv:1107.1452 [hep-ph]}
  \BibitemShut {NoStop}%
\bibitem [{\citenamefont {Kopp}\ \emph {et~al.}(2011)\citenamefont {Kopp},
  \citenamefont {Maltoni},\ and\ \citenamefont {Schwetz}}]{Kopp:2011qd}%
  \BibitemOpen
  \bibfield  {author} {\bibinfo {author} {\bibfnamefont {J.}~\bibnamefont
  {Kopp}}, \bibinfo {author} {\bibfnamefont {M.}~\bibnamefont {Maltoni}}, \
  and\ \bibinfo {author} {\bibfnamefont {T.}~\bibnamefont {Schwetz}},\ }\href
  {\doibase 10.1103/PhysRevLett.107.091801} {\bibfield  {journal} {\bibinfo
  {journal} {Phys. Rev. Lett.}\ }\textbf {\bibinfo {volume} {107}},\ \bibinfo
  {pages} {091801} (\bibinfo {year} {2011})},\ \Eprint
  {http://arxiv.org/abs/1103.4570} {arXiv:1103.4570 [hep-ph]} \BibitemShut
  {NoStop}%
\bibitem [{\citenamefont {Komatsu}\ \emph {et~al.}(2011)\citenamefont {Komatsu}
  \emph {et~al.}}]{Komatsu:2010fb}%
  \BibitemOpen
  \bibfield  {author} {\bibinfo {author} {\bibfnamefont {E.}~\bibnamefont
  {Komatsu}} \emph {et~al.} (\bibinfo {collaboration} {WMAP}),\ }\href
  {\doibase 10.1088/0067-0049/192/2/18} {\bibfield  {journal} {\bibinfo
  {journal} {Astrophys. J. Suppl.}\ }\textbf {\bibinfo {volume} {192}},\
  \bibinfo {pages} {18} (\bibinfo {year} {2011})},\ \Eprint
  {http://arxiv.org/abs/1001.4538} {arXiv:1001.4538 [astro-ph.CO]} \BibitemShut
  {NoStop}%
\bibitem [{c:f(2005)}]{c:ffag}%
  \BibitemOpen
  \href@noop {} {}\bibinfo {howpublished}
  {\url{http://hadron.kek.jp/FFAG/FFAG05_HP/index.htm}} (\bibinfo {year}
  {2005}),\ \bibinfo {note} {see talks including the summary talk}\BibitemShut
  {NoStop}%
\bibitem [{\citenamefont {Huber}\ \emph {et~al.}(2005)\citenamefont {Huber},
  \citenamefont {Lindner},\ and\ \citenamefont {Winter}}]{Huber:2004ka}%
  \BibitemOpen
  \bibfield  {author} {\bibinfo {author} {\bibfnamefont {P.}~\bibnamefont
  {Huber}}, \bibinfo {author} {\bibfnamefont {M.}~\bibnamefont {Lindner}}, \
  and\ \bibinfo {author} {\bibfnamefont {W.}~\bibnamefont {Winter}},\ }\href
  {\doibase 10.1016/j.cpc.2005.01.003} {\bibfield  {journal} {\bibinfo
  {journal} {Comput.Phys.Commun.}\ }\textbf {\bibinfo {volume} {167}},\
  \bibinfo {pages} {195} (\bibinfo {year} {2005})},\ \Eprint
  {http://arxiv.org/abs/hep-ph/0407333} {arXiv:hep-ph/0407333 [hep-ph]}
  \BibitemShut {NoStop}%
\bibitem [{\citenamefont {Huber}\ \emph {et~al.}(2007)\citenamefont {Huber},
  \citenamefont {Kopp}, \citenamefont {Lindner}, \citenamefont {Rolinec},\ and\
  \citenamefont {Winter}}]{Huber:2007ji}%
  \BibitemOpen
  \bibfield  {author} {\bibinfo {author} {\bibfnamefont {P.}~\bibnamefont
  {Huber}}, \bibinfo {author} {\bibfnamefont {J.}~\bibnamefont {Kopp}},
  \bibinfo {author} {\bibfnamefont {M.}~\bibnamefont {Lindner}}, \bibinfo
  {author} {\bibfnamefont {M.}~\bibnamefont {Rolinec}}, \ and\ \bibinfo
  {author} {\bibfnamefont {W.}~\bibnamefont {Winter}},\ }\href {\doibase
  10.1016/j.cpc.2007.05.004} {\bibfield  {journal} {\bibinfo  {journal}
  {Comput.Phys.Commun.}\ }\textbf {\bibinfo {volume} {177}},\ \bibinfo {pages}
  {432} (\bibinfo {year} {2007})},\ \Eprint
  {http://arxiv.org/abs/hep-ph/0701187} {arXiv:hep-ph/0701187 [hep-ph]}
  \BibitemShut {NoStop}%
\bibitem [{\citenamefont {Kopp}(2008)}]{Kopp:2006wp}%
  \BibitemOpen
  \bibfield  {author} {\bibinfo {author} {\bibfnamefont {J.}~\bibnamefont
  {Kopp}},\ }\href {\doibase 10.1142/S0129183108012303} {\bibfield  {journal}
  {\bibinfo  {journal} {Int. J. Mod. Phys.}\ }\textbf {\bibinfo {volume}
  {C19}},\ \bibinfo {pages} {523} (\bibinfo {year} {2008})},\ \bibinfo {note}
  {erratum ibid.\ {\bf C19} (2008) 845},\ \Eprint
  {http://arxiv.org/abs/physics/0610206} {arXiv:physics/0610206} \BibitemShut
  {NoStop}%
\bibitem [{\citenamefont {Kopp}\ \emph {et~al.}(2008)\citenamefont {Kopp},
  \citenamefont {Lindner}, \citenamefont {Ota},\ and\ \citenamefont
  {Sato}}]{Kopp:2007ne}%
  \BibitemOpen
  \bibfield  {author} {\bibinfo {author} {\bibfnamefont {J.}~\bibnamefont
  {Kopp}}, \bibinfo {author} {\bibfnamefont {M.}~\bibnamefont {Lindner}},
  \bibinfo {author} {\bibfnamefont {T.}~\bibnamefont {Ota}}, \ and\ \bibinfo
  {author} {\bibfnamefont {J.}~\bibnamefont {Sato}},\ }\href {\doibase
  10.1103/PhysRevD.77.013007} {\bibfield  {journal} {\bibinfo  {journal} {Phys.
  Rev.}\ }\textbf {\bibinfo {volume} {D77}},\ \bibinfo {pages} {013007}
  (\bibinfo {year} {2008})},\ \Eprint {http://arxiv.org/abs/0708.0152}
  {arXiv:0708.0152 [hep-ph]} \BibitemShut {NoStop}%
\bibitem [{\citenamefont {Tunnell}(2011)}]{vlenf_tools}%
  \BibitemOpen
  \bibfield  {author} {\bibinfo {author} {\bibfnamefont {C.~D.}\ \bibnamefont
  {Tunnell}},\ }\href@noop {} {}\bibinfo {howpublished}
  {\url{https://code.launchpad.net/~c-tunnell1/+junk/vlenf_tools}} (\bibinfo
  {year} {2011}),\ \bibinfo {note} {questions should be directed to the
  corresponding author.}\BibitemShut {Stop}%
\bibitem [{gpl()}]{gpl}%
  \BibitemOpen
  \href@noop {} {}\bibinfo {howpublished}
  {\url{http://www.gnu.org/licenses/gpl-3.0.html}}\BibitemShut {NoStop}%
\bibitem [{\citenamefont {Laing}(2010)}]{laing}%
  \BibitemOpen
  \bibfield  {author} {\bibinfo {author} {\bibfnamefont {A.~B.}\ \bibnamefont
  {Laing}},\ }\href@noop {} {\enquote {\bibinfo {title} {Optimisation of
  detectors for the golden channel at a neutrino factory},}\ } (\bibinfo {year}
  {2010})\BibitemShut {NoStop}%
\bibitem [{\citenamefont {Giunti}\ and\ \citenamefont
  {Laveder}(2011{\natexlab{b}})}]{Giunti:2011hn}%
  \BibitemOpen
  \bibfield  {author} {\bibinfo {author} {\bibfnamefont {C.}~\bibnamefont
  {Giunti}}\ and\ \bibinfo {author} {\bibfnamefont {M.}~\bibnamefont
  {Laveder}},\ }\href@noop {} {\  (\bibinfo {year} {2011}{\natexlab{b}})},\
  \Eprint {http://arxiv.org/abs/1109.4033} {arXiv:1109.4033 [hep-ph]}
  \BibitemShut {NoStop}%
\end{thebibliography}%

\end{document}